\documentclass[a4paper]{article}

\usepackage{authblk}
\usepackage{graphicx}
\usepackage{natbib}
\usepackage{amssymb,amsmath,color,epic,eepic,graphicx}
\usepackage{MnSymbol}

\title{\textbf{Inertial particle acceleration in strained turbulence}}

\author[1,4]{\normalsize Chung-min Lee}
\author[2]{\normalsize \'{A}rmann Gylfason\thanks{Email address for correspondence: armann@ru.is}}
\author[3]{\normalsize Prasad Perlekar}
\author[4,5,6]{\normalsize Federico Toschi}

\affil[1]{\footnotesize Department of Mathematics and Statistics, California State University Long Beach, USA}
\affil[2]{\footnotesize School of Science and Engineering, Reykjav\'ik University, Menntavegur 1, 101 Iceland}
\affil[3]{\footnotesize TIFR Centre for Interdisciplinary Sciences, Hyderabad 500075, India}
\affil[4]{\footnotesize Department of Applied Physics, Eindhoven University of Technology, The Netherlands}
\affil[5]{\footnotesize Department of Mathematics and Computer Science, Eindhoven University of Technology, The Netherlands}
\affil[6]{\footnotesize Istituto per le Applicazioni del Calcolo CNR, Via dei Taurini 19, 00185 Rome, Italy}

\date{}

\begin{document}

\maketitle

\begin{abstract}
The dynamics of inertial particles in turbulence is modelled and investigated by means of direct
numerical simulation of an axisymmetrically expanding homogeneous
turbulent strained flow.  This flow can mimic the dynamics of particles close to stagnation points.  The influence of mean straining flow is
explored by varying the dimensionless strain rate parameter
$Sk_0/\epsilon_0$ from 0.2 to 20. We report results relative to the
acceleration variances and probability density functions for both
passive and inertial particles. A high mean strain is found to have a
significant effect on the acceleration variance both directly,
through an increase in wave number magnitude, and indirectly, through
the coupling of the fluctuating velocity and the mean flow field. The
influence of the strain on normalized particle acceleration pdfs is
more subtle.  For the case of passive particle we can approximate the
acceleration variance with the aid of rapid distortion theory and
obtain good agreement with simulation data. For the case of inertial
particles we can write a formal expressions for the accelerations. The
magnitude changes in the inertial particle acceleration variance and
the effect on the probability density function are then discussed in
a wider context for comparable flows, where the effects of the mean
flow geometry and of the anisotropy at the small scales are present.
\end{abstract}

\section{Introduction}
The motions of small, passive and inertial particles in turbulence has
been extensively studied in recent years, both from the experimental
and theoretical viewpoints. This was motivated by a broad range of
applications such as the spread of pollutants in the atmosphere and
oceans, the process of rain and ice formation in cloud, the transport
of sediments in rivers and estuaries, see e.g. the reviews of
\cite{S2003} and \cite{TB2009}. Progress in the understanding was on the one hand made
possible by recent improvements in Lagrangian measurements through particle tracking methodologies, resulting in part from rapid advances in high speed imaging (\cite{VD,OM,VPCAB,XOB_2008}), and on the other hand by increased computational capabilities of numerical simulations (\cite{YP_1998,C_2007}).

The objective of this work is to investigate the effects of flow
straining on the Lagrangian dynamics of small, sub-Kolmogorov
scale, passive and inertial particles. Our motivation stems both from
the fact that many practical turbulent flows are subject to straining
motions, such as the external flows over bluff or streamlined bodies
and internal flows in variable cross sections (\cite{B_1953, H1973,
  HC, W_1980, CMK, AW, GM_2010}). A mean straining flow naturally appears in the
proximity of stagnation points. Flow straining is furthermore of
fundamental interest since it induces a scale dependent anisotropy;
the smallest scales of the flow may be nearly isotropic, whereas the
largest scales are highly anisotropic (\cite{BP_2005}).

Furthermore, many flows naturally combine straining geometries and inertial
particles.
The flow geometry presented here, namely particle laden turbulent flow
undergoing an axisymmetric expansion, has similarities with combustor diffusers in jet engines (\cite{K_1995}), where liquid fuel is injected in an expanding flow, and with the flow in the combustion chamber in an internal combustion engine during the compression stroke of the fuel air mixture (\cite{HR_1995}).

While significant attention has been given to the study of Lagrangian
acceleration statistics in isotropic turbulence, less attention has
been paid to the implications of anisotropic large scale flow geometry
on the Lagrangian dynamics. Recent experimental and numerical work on
the Lagrangian behavior of inertial particles in shear flows and in turbulent boundary
layer has shown pronounced effects on the inertial particle statistics 
(\cite{GSNW_2008,LSGWC2010, GPC_2009, GPSC_2012}). The persistent small scale anisotropy has been found to influence the geometry and alignment of particle clusters and relative particle pair velocities. In addition, the combined effects of gravity and shear on particle acceleration variance results in an increase in magnitude with
the Stokes number. As a consequence, the acceleration probability
distribution functions (pdfs) became increasingly narrow and skewed
with inertia. Here, we address a related topic, namely the complexity
introduced in the Lagrangian dynamics of tracer and inertial particles
due to flow straining. 

In an effort to realize the effects of anisotropy in the particle
dynamics, we numerically simulate axisymmetric expansion of initially
isotropic turbulence. The flow is seeded with infinitesimal tracer and
inertial particles of varied Stokes numbers. We measure the particle
velocity and acceleration statistics, including variances and
probability density functions for different strain rates and Stokes
numbers. Comparisons are made with predictions of rapid distortion
theory on tracer accelerations, and the solutions of the Stokes
equations for inertial particles in the straining flow.

The paper is organized as follows.  In section 2 we briefly introduce
the numerical methods for simulating an axisymmetric turbulence and
particle movements.  Parameters of the simulations are listed.
Section 3 presents the underlying flow field.  We discuss our main
findings of particle acceleration variances and probability density
functions in simulation data with the support of theoretical
estimations in section 4.  We also discuss our result in the context
of previous work in shear flows.  In section 5 we present our
conclusions.

\section{Methodology}
\subsection{Flow equations and flow simulation}
\label{flowfieldsim}

The equations describing the motion of an incompressible Newtonian
fluid are the continuity equation and the Navier-Stokes equations,
respectively:
\begin{eqnarray}
\nabla \cdot \tilde{\boldsymbol{ u}} &=& 0, \nonumber
\\ \frac{\partial \tilde{\boldsymbol{ u}}}{\partial t} +
\tilde{\boldsymbol{ u}} \cdot \nabla \tilde{\boldsymbol{u}} + \nabla
\tilde{p} &=& \nu \nabla^2 \tilde{\boldsymbol{ u}}.
\label{NSeincomp}
\end{eqnarray}
Here, $\tilde{\boldsymbol{ u}}$ and $\tilde{p}$ are the instantaneous flow velocity and pressure, respectively, and $\nu$ is the kinematic viscosity of the fluid.  

\begin{figure}
\vspace*{2mm}
\begin{center}
\scalebox{0.35}{\input{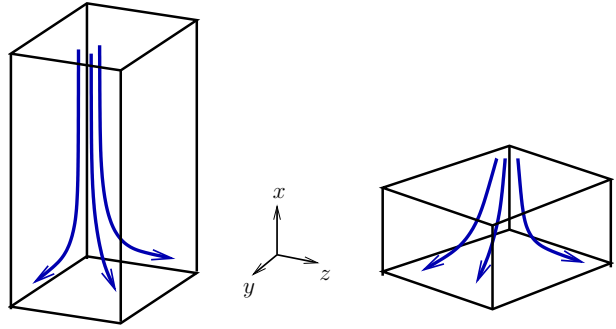}}
\caption{A sketch of the  deformation of the simulation domain under straining. The mean flow $\boldsymbol{U} = (-2Sx, Sy, Sz)$ corresponds to an ideal flow onto a flat plate. The deforming domain is initially elongated in the $x$-direction but becomes wider in $y$ and $z$ directions with time.   Arrows indicate the directions of the stream lines of the induced mean flow.
}
\label{Sketch_Expansion}
\end{center}
\end{figure}

In this paper we are concerned with a turbulent fluid undergoing an axisymmetric expansion, where the mean flow field is described by 
\begin{equation}
\label{meanflow}
\boldsymbol{U} = (-2Sx, Sy, Sz), 
\end{equation}
here $S$ is the mean strain rate $S=\frac1{\sqrt{6}}\left(\bar{S}_{ij}\bar{S}_{ij}\right)^{1/2}$, and $\bar{S}_{ij} = \frac12\left(\frac{\partial U_i}{\partial x_j}+\frac{\partial U_j}{\partial x_i}\right)$ is the mean rate of strain tensor. The mean flow corresponds to an ideal flow onto a flat plate, and is realized in the flow between contracting pistons or in the expanding flow through a diffuser. Figure \ref{Sketch_Expansion} shows a sketch of the mean flow field, namely streamlines of the mean field, and the deforming domain. 

Applying the Reynolds decomposition, and expressing (\ref{NSeincomp}) in terms of the vector potential $\boldsymbol{b}$, with $\boldsymbol{u} = \nabla \times \boldsymbol{b}$, one obtains 
\begin{equation}
\label{vecpotentiale}
-{\partial_t \nabla^2 \boldsymbol{b}} - \nabla \times ({\boldsymbol u} \times {\boldsymbol \omega}) + 2Sx \partial_x \nabla^2 {\boldsymbol b} - Sy \partial_y \nabla^2 {\boldsymbol b} - Sz \partial_z \nabla^2 {\boldsymbol b} + S \nabla^2 {\boldsymbol b} - 3S \nabla^2 b_1 \hat{e}_1 = -\nu \nabla^4 {\boldsymbol b},
\end{equation}
where $\boldsymbol u = \tilde{\boldsymbol{ u}} - \boldsymbol{U}$ is the velocity fluctuation, $\boldsymbol{ \omega}$ is the vorticity, defined as $\boldsymbol{ \omega} \equiv \nabla \times {\boldsymbol u}=-\nabla^2 \boldsymbol{b}$. $b_1$ is the first component of ${\boldsymbol b}$ and $\hat{e}_1$ is the unit vector in the $x$-direction.  In the following we briefly outline the numerical algorithm.

As in \cite{GLPT_2011}, in order to solve (\ref{vecpotentiale}) we apply a pseudo-spectral method with Rogallo's algorithm (\cite{Rogallo_81}) where the following variable transformations are performed:
\[
x' = e^{2St}x, \hspace{1cm} y' = e^{-St}y, \hspace{1cm} z' = e^{-St}z,\hspace{1cm} t'= t,
\]
and hence the vector potential of the velocity fluctuation satisfies
\begin{equation}
\label{vecpotentialprime}
{\partial_{t'} \nabla^{'2} {\boldsymbol b}} - \nabla' \times ({\boldsymbol u} \times {\boldsymbol \omega}) + S \nabla^{'2} {\boldsymbol b} -3S \nabla^{'2} b_1 \hat{e}_1 = -\nu \nabla^{'4} {\boldsymbol b},
\end{equation}
where $\nabla' = \left( e^{2St} \partial_{x'},  e^{-St} \partial_{y'},  e^{-St} \partial_{z'} \right)$.
By adopting this new coordinate system, the physical domain deforms with time while the computational lattice grid is time independent, and the flow equations become periodic. We then apply the pseudo-spectral method to equations in (\ref{vecpotentialprime}).  More details about the numerical method can be found in \cite{GLPT_2011}.

Numerical simulations of this axisymmetric expansion flow were carried out on a Cartesian grid with 1024$\times$256$\times$256 and 2048$\times$512$\times$512 grid points in the $x$, $y$ and $z$ directions, respectively.  The initial configurations are derived from statistically independent homogeneous and isotropic flow simulations which have reached a stationary state after more than 5 large-eddy turnover times.  The Reynolds numbers, based on the Taylor microscale $\lambda_0$ were $R_{\lambda0} = 117$ and $R_{\lambda0} = 193$ before the straining was applied, for the lower and higher grid resolution respectively. Initially the physical domain size was $[0,8\pi]\times[0,2\pi]\times[0,2\pi]$ in $x$, $y$ and $z$ directions respectively, and the simulation was terminated when the domain has reached  $[0,1.1\pi]\times[0,5.4\pi]\times[0,5.4\pi]$ to prevent the physical domain to become too flattened. 

\begin{figure}
\vspace*{4mm}
\begin{center}
\hspace*{-12mm}
\includegraphics[width=6.1in]{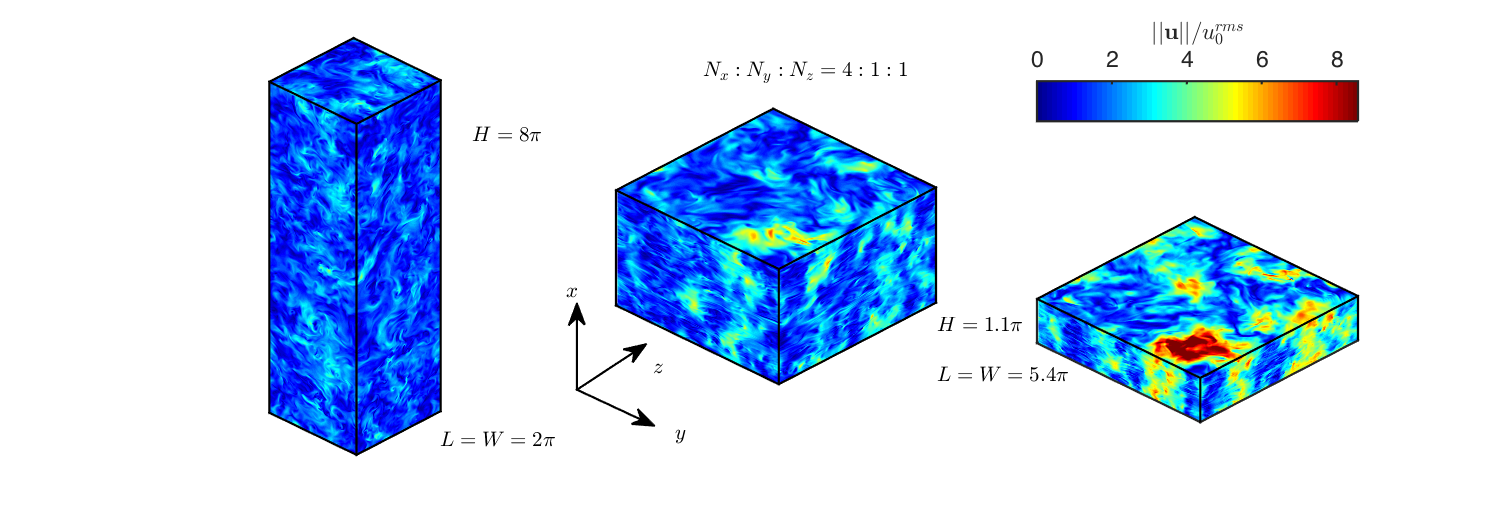}\\
\vspace*{7mm}
\hspace*{-12mm}
\includegraphics[width=6.1in]{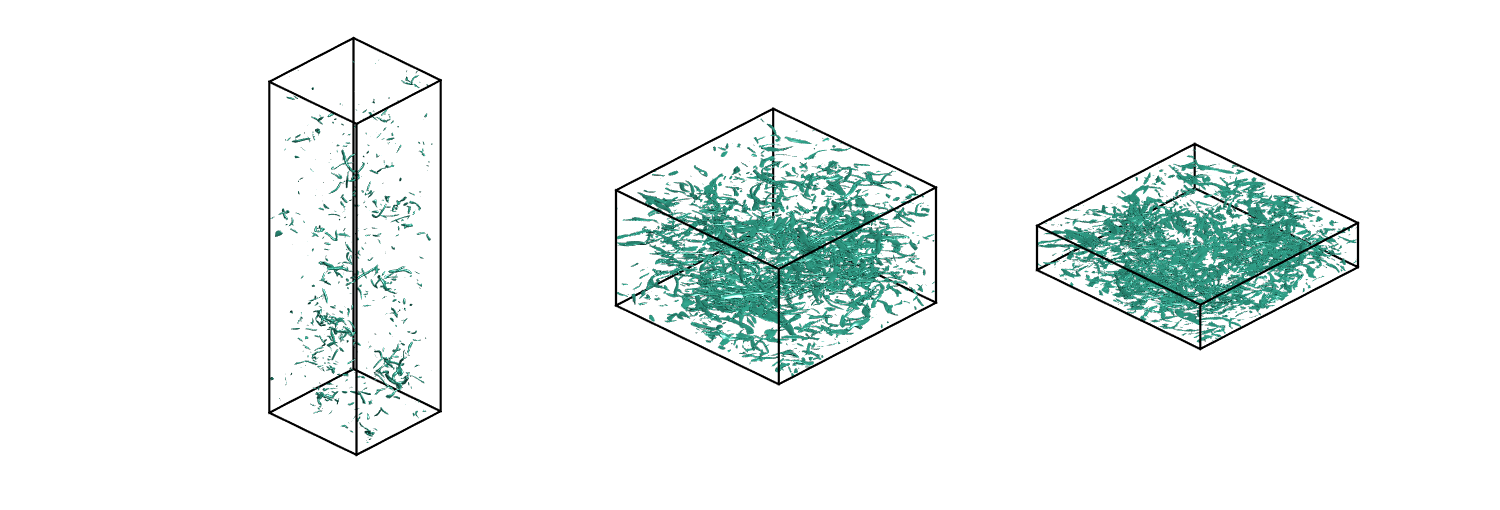}
\caption{Snapshots of the magnitude of the fluctuating flow velocity (top) and isosurface of the magnitude of vorticity $\omega=||\nabla \times{\boldsymbol u}||$  (bottom) in the deforming domain in a realization of the axisymmetrically expanding flow. Left to right: from the onset of the straining simulation to the end of the straining simulation at time instants $S\times t=0.08, 0.64$, and $0.96$. The size of the simulation domain is noted in the figure, and its deformation is displayed. The coordinate system adopted in the text is also shown.  We simulated the axisymmetric turbulence with two resolutions using $2048\times 512\times 512$ and $1024\times 256\times 256$ computational nodes.  The illustration shows one of the realizations of the $1024\times 256\times 256$ simulations with $S=10$.  The isosurfaces plotted have non-dimensional vorticity $\omega/(\varepsilon_0/\nu)^{1/2} = 3.17$, which is respectively 4.36, 2.86 and 2.43 standard deviations above the mean vorticity magnitude at the three time instants and chosen to illustrate the flow structure.}
\label{velmag}
\end{center}
\end{figure}

The top of Figure \ref{velmag} shows snapshots of the fluctuating velocity magnitude at three time instants during the straining. From left to right, the non-dimensional times are $S\times t=0.08$ (shortly after the mean strain is applied), at $S\times t=0.64$, and at $S\times t=0.96$ (just before the strain simulation is terminated due the large deformation of the physical domain). Additionally, the figure shows the coordinate system adopted in the text, and the geometry of the simulation domain selected and its deformation. Production of turbulence overwhelms dissipation during the straining, reflected in an increase in the turbulent kinetic energy, most notably in the compressed component ($x$).  This can be seen in the warmer colors in the rightmost plot. 
The bottom of Figure \ref{velmag} shows isosurfaces of non-dimensional vorticity $\omega/(\varepsilon_0/\nu)^{1/2} = 3.17$ at the same time instants as above, which is respectively 4.36, 2.86 and 2.43 standard deviations above the mean vorticity magnitude at the three time instants. From the increased number and size of the filaments, we observe that the vorticity is intensified during straining, and the filaments are found to gradually align with the $y,z$-plane due to the mean flow extension in the plane.

\subsection{Equations for particle movements}
\label{particlesim}
To study Lagrangian aspects of this flow we seed the flow with tracers and inertial particles. Here, we are concerned with particles that are small compared to the smallest length scales present in the flow and their densities considerably higher than the fluid density. The particle number densities are furthermore assumed to be sufficiently low so that particle-particle interactions can be ignored. Under the above approximations, the coupling of the particles with the carrier fluid can be ignored.

The Lagrangian equations of inertial particle motion is derived from Newton's second law, and represents the balance between the forces acting on the particles (inertia and Stokes drag). The equations describing the motion of a particle of diameter $d_p$ and density $\rho_p$, located at $\boldsymbol{x}_p$ and with instantaneous velocity $\tilde{\boldsymbol{v}}_p$ are (\cite{MR, BBBCCLMT}):
\begin{eqnarray}
\frac{d \boldsymbol{x}_p}{dt} &=& \tilde{\boldsymbol{v}}_p, \label{particleloc}\\
\frac{d \tilde{\boldsymbol{v}}_p}{dt} &=& \frac1{\tau_p}\left(\tilde{\boldsymbol{u}}(\boldsymbol{x}_p) - \tilde{\boldsymbol{v}}_p\right), 
\label{particlevel}
\end{eqnarray}
where $\tau_p = \beta d_p^2 / 18\nu$ is the Stokes relaxation time for the particle and $\beta=(\rho_p - \rho_f)/\rho_f$ is the relative density ratio between the particle and the fluid. The Stokes number $St = \tau_p/\tau_\eta$ characterizes the inertia of a particle in the flow, where $\tau_\eta$ is the Kolmogorov timescale of the flow.

For the tracer particles (zero inertia), the particle velocity is the same as the fluid velocity at the particle location, which yields the kinematic relation:
\begin{equation}
\frac{d \boldsymbol{x}_p}{dt} = \tilde{\boldsymbol{u}}(\boldsymbol{x}_p). \label{tracerparticleloc}
\end{equation}

The ordinary differential equations (\ref{particleloc}), (\ref{particlevel}), and  (\ref{tracerparticleloc}) are solved numerically by the second order Adams-Bashforth method. 
In equations (\ref{particlevel}) and (\ref{tracerparticleloc}), the instantaneous flow velocity at the particle location, $\boldsymbol{x}_p$, is evaluated as
\[
\tilde{\boldsymbol{u}}(\boldsymbol{x}_p) = \boldsymbol{U}(\boldsymbol{x}_p) + \boldsymbol{u}(\boldsymbol{x}_p);
\]
that is, the mean flow velocity is evaluated at the location of the particle through the formula $\boldsymbol{U}(\boldsymbol{x}_p) = (-2Sx_p, Sy_p, Sz_p)$, and the flow velocity fluctuation is interpolated to the particle position.

We initialize the particles and the fluid velocity with steady state homogeneous isotropic simulations. The particles are uniformly distributed over the domain prior to the forced homogeneous isotropic simulation is carried out. At the beginning of the straining, at $t=0$, we add the mean flow velocity and acceleration component due to the strain geometry to the existing particle velocity. We conduct simulations with 1024$\times$256$^2$ and 2048$\times$512$^2$ collocation points. For the lower resolution simulation we use 16 independent flow realizations with 5$\times$10$^5$ particles of each type (6 different Stokes numbers) and for the higher resolution simulations we perform 10 independent flow realizations with 4$\times$10$^5$ particles of each type.

Table \ref{table1} shows the various flow parameters for the simulations performed. The range of strain rates selected is such that its effect on the smallest scales of the flow range from being negligible to substantial. The higher strain rates are felt intensely by the large scale flow, whereas the lower strain rates have mild effect on the large scales. The value of the strain parameters, $S\tau_{\eta0}$ and $Sk_0/\epsilon_0$, which compare the strain time with the local timescales of the flow, indicate the importance of the various terms in the evolution equation of the velocity field. 
\begin{table}
\caption{Flow parameters in the Direct Numerical Simulations, based on the homogeneous isotropic simulation prior to the application of the straining.  Here $k_0 \equiv \frac12\left(\langle u_{10}^2\rangle+\langle u_{20}^2\rangle+\langle u_{30}^2\rangle\right)$ is the turbulent kinetic energy, $\epsilon_0 \equiv \frac12\langle\frac{\partial u_{i0}}{\partial x_j}\frac{\partial u_{i0}}{\partial x_j}\rangle$ is the energy dissipation rate, $\ell_0 \equiv (u_0^{rms})^3/\epsilon_0$ is the integral length scale, and the Kolmogorov length scale $\eta_0 \equiv \left(\nu^3/\epsilon_0\right)^{1/4}$.  The subscript $0$ indicates the parameter values are taken prior to the straining.  (Units are arbitrary)}
\label{tab_flowpar}
\begin{center}
\begin{tabular}{lcc}
\hline
Simulation domain & 1024$\times$256$\times$256 & 2048$\times$512$\times$512 \\
$R_{\lambda0} \equiv u_0^{rms}\lambda_0/\nu$ & 117 & 193\\
$k_0 \equiv \frac12\left(\langle u_{10}^2\rangle+\langle u_{20}^2\rangle+\langle u_{30}^2\rangle\right)$ & 4.6 & 4.9\\
$\lambda_0/\eta_0 \equiv u_0^{rms}(15\nu/\epsilon_0)^{1/2}/\eta_0$ & 20.9 & 27.4\\
$S$ & $0.1, 0.5, 1, 4, 10$ &$1, 4, 10$ \\ 
$\tau_{\eta0} \equiv (\nu/\epsilon_0)^{1/2}$ & 0.051 & 0.031 \\
$u_0^{rms} = (2k_0/3)^{1/2}$ & 1.75 & 1.81\\
$\ell_0/\eta_0 \equiv (u_0^{rms})^3/\epsilon_0/\eta_0$ &  164.4& 332.6 \\
$\eta_0 \equiv (\nu^3/\epsilon_0)^{1/4}$ & 0.0163 $\pm$  0.0006 & 0.008 $\pm$  0.0004\\
$\epsilon_0 \equiv \frac12\langle\frac{\partial u_{i0}}{\partial x_j}\frac{\partial u_{i0}}{\partial x_j}\rangle$ & 2.18 $\pm$  0.15 & 2.12 $\pm$ 0.4 \\
$\nu$ & 0.0052 & 0.00205\\
$St_0=\tau_p/\tau_{\eta0}$ & $0, 0.2, 0.3, 0.5, 1, 2$& $0, 0.23, 0.34, 0.56, 1.12, 2.25$\\
$S\tau_{\eta0}$ & $0.0051, 0.0255, 0.051, 0.204, 0.51$ & $0.031, 0.124, 0.31$\\
$Sk_0/\epsilon_0$ & $0.21, 1.06, 2.11, 8.44, 21.1$&$2.31, 9.25, 23.1$\\
\hline
\label{table1}
\end{tabular}
\end{center}
\end{table}

\section{Underlying flow field}

\begin{figure}
\vspace*{5mm}
\begin{center}
\includegraphics[height=2.5in,width=5in]{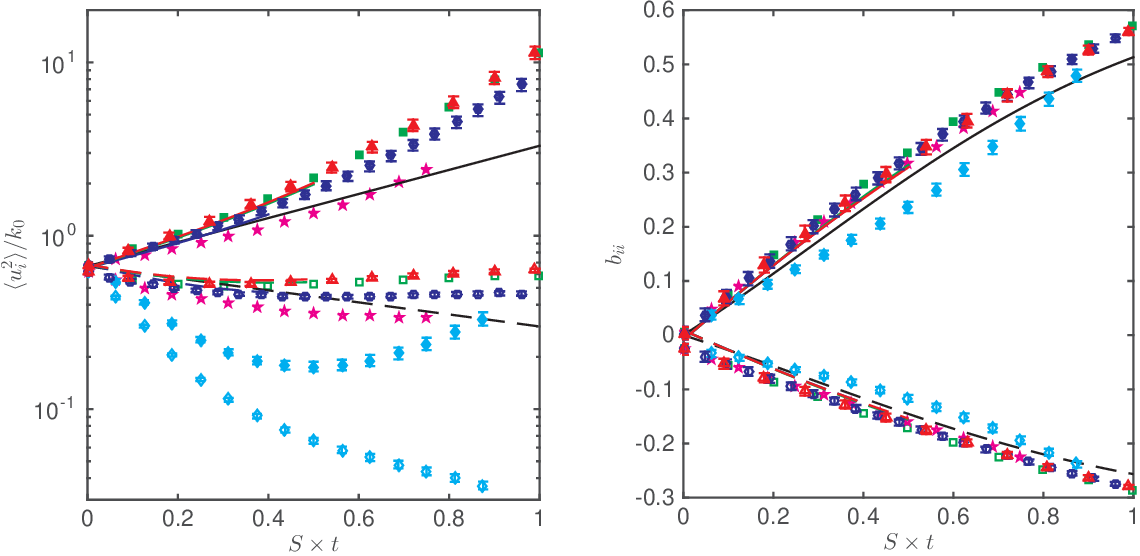}
\caption{Left: Normalized Reynolds stresses $\langle u_i^2 \rangle / k_0$ vs. strain time $S\times t$. Right: The flow anisotropy tensor $\displaystyle b_{ij} = \frac{\langle u_iu_j \rangle}{\langle u_iu_i \rangle} - \frac{1}{3}\delta_{ij}$ vs. strain time. Solid symbols represent the $i=1$ component and open symbols represent the $i=2$ component.  $k_0$ is the initial kinetic energy of the fluid prior to the straining.  The  diamond ($\meddiamond$), pentagram ($\medstar$), circle ($\bigcircle$), square ($\medsquare$) and triangle ($\medtriangleup$) symbols represent data from $S=0.1, 0.5, 1, 4, 10$ in the $R_{\lambda0} = 117$ flow, respectively.  Blue, green and red lines, indicate the data from $S=1, 4, 10$ in the $R_{\lambda0} = 193$ flow; solid and dashed lines represent the $i=1$ and $i=2$ components.  Black solid and dashed lines represent the short term RDT predictions (\ref{RDTearlyvel}).  Estimates of statistical error bars on data from $S=0.1,1, 10$ are shown and are computed according to  (\ref{stderr}) with $X_j$ being $\langle u_i^2 \rangle / k_0$ and $b_{ii}$ in the $j$-th realization.
}
\label{RstressAniso}
\end{center}
\end{figure}

The left hand side of Figure \ref{RstressAniso} shows the evolution of the Reynolds stresses ($\langle u_i^2 \rangle$) normalized by the initial turbulent kinetic energy ($k_0$).  The component along the compressed direction ($x_1$) grows rapidly in most cases, whereas the components along the expanding directions are suppressed or remain roughly constant. At the lowest strain rate all component are suppressed during the simulation time. For large times, rapid distortion theory (RDT) predicts an exponential growth of the Reynolds stresses, in the proportions $\langle u_1^2\rangle = 2\langle u_2^2\rangle = 2\langle u_3^2\rangle$. The kinetic energy increases with time for all the strain rates, although the lower rates display an initial drop in energy and a subsequent long term increase.  The right hand side of the figure shows the evolution of the anisotropy tensor $b_{ij} = \frac{\langle u_iu_j \rangle}{\langle u_iu_i \rangle} - \frac{1}{3}\delta_{ij}$ with time. The curves corresponding to the lowest strain rates markedly deviate from the others as the turbulent kinetic energy decreases during the straining, and the straining motions are fairly mild, even for the largest scales of motions.

The short term RDT prediction plotted in Figure \ref{RstressAniso} is derived from the Reynolds stress equation (\cite{P_2000}) 
\begin{equation}
\frac{d}{dt}\langle u_i u_j \rangle = \mathcal P_{ij} + \mathcal R_{ij}^{(r)},
\end{equation}
where $\mathcal P_{ij} \equiv -\langle u_i u_k \rangle \frac{\partial U_j}{\partial x_k} - \langle u_j u_k \rangle \frac{\partial U_i}{\partial x_k}$ is the production rate of Reynolds stress, $\mathcal R_{ij}^{(r)} \equiv \langle \frac{p^{(r)}}{\rho}\left(\frac{\partial u_i}{\partial x_j} + \frac{\partial u_j}{\partial x_i}\right)\rangle$ is the rapid pressure-rate-of-strain tensor, and $p^{(r)}$ is the rapid pressure that satisfies $\frac1{\rho}\nabla^2 p^{(r)} = -2\frac{\partial U_i}{\partial x_j}\frac{\partial u_j}{\partial x_i}$.  Right before the straining starts, the initial configuration of flow is isotropic, and $\mathcal R_{ij}^{(r)} = -\frac35 \mathcal P_{ij}$ (\cite{P_2000}).  In an axisymmetric expansion flow, the production rates are $\mathcal P_{11} = 4S\langle (u_1)^2 \rangle, \mathcal P_{22} = -2S\langle (u_2)^2 \rangle$, and $\mathcal P_{33} = -2S\langle (u_3)^2 \rangle$.  Therefore, in early times RDT predicts
\begin{equation}
\langle (u_1)^2 \rangle = \langle (u_{10})^2 \rangle e^{\frac85 St},\ \langle (u_2)^2 \rangle = \langle (u_{20})^2 \rangle e^{-\frac45 St},\ \langle (u_3)^2 \rangle = \langle (u_{30})^2 \rangle e^{-\frac45 St},
\label{RDTearlyvel}
\end{equation}
where $\langle (u_{i0})^2 \rangle$ represent the initial values of the Reynolds stresses ($\langle (u_{i})^2 \rangle, i = 1, 2, 3$). However, in order for rapid distortion theory to apply, the parameters must satisfy $S\tau_\eta \gg 1$ and $Sk/\epsilon \gg 1$ (\cite{B_1953}). Only at the highest rate of strain is the latter constraint weakly satisfied, and therefore one does not observe close matches between the predictions of RDT and the Reynolds stresses in simulation data.  The global anisotropy is much less sensitive, and short term RDT predicts the anisotropy well.   

\begin{figure}
\vspace*{2mm}
\begin{center}
\includegraphics[height=2.4in,width=3.2in]{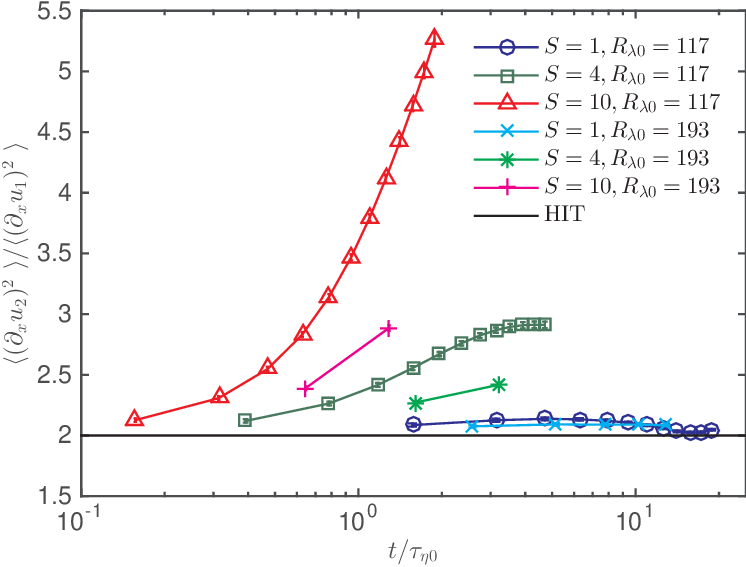}
\caption{Ratios of the variances of velocity derivatives  $\langle (\partial_x u_2)^2\rangle/\langle(\partial_x u_1)^2\rangle$ vs. normalized time $t / \tau_{\eta0}$.  The circle ($\bigcircle$), square ($\medsquare$) and triangle ($\medtriangleup$) symbols represent strain rates $S=1, 4, 10$ from the $R_{\lambda0} = 117$ set, and crosses ($\times$), asterisks (\textasteriskcentered) and pluses ($+$) represent strain rates  $S=1, 4, 10$ from the $R_{\lambda0} = 193$ set.  An estimate of statistical error bar is shown (within the symbols) and computed according to   (\ref{stderr}) with $X_j$ being $\langle (\partial_x u_2)^2\rangle/\langle(\partial_x u_1)^2\rangle$ in the $j$-th realization. The solid line shows the theoretical prediction for this ratio in the isotropic turbulence.}
\label{du2dv2}
\end{center}
\end{figure}

\begin{figure}
\vspace*{4mm}
\begin{center}
\includegraphics[width=2.52in]{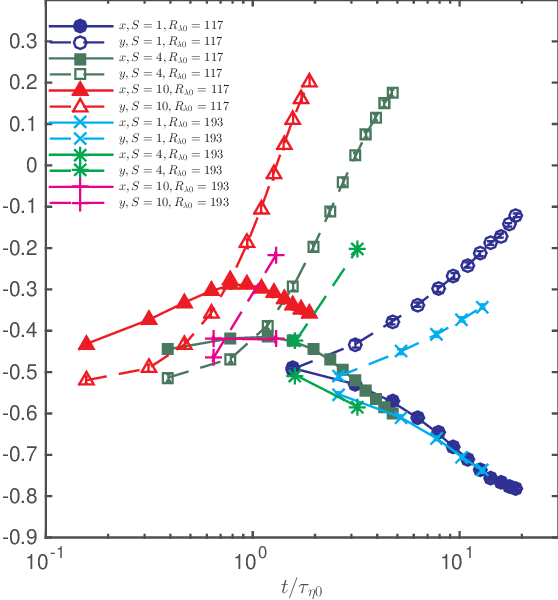}
\hspace*{4mm}
\includegraphics[width=2.45in]{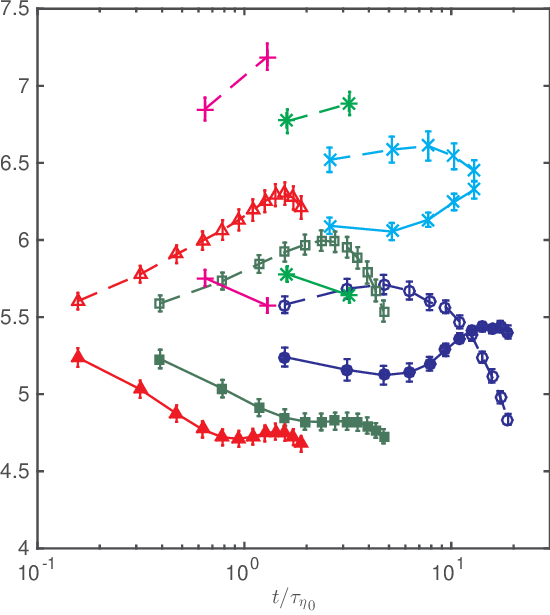}
\end{center}
\caption{Left: Skewness in the compression and expansion directions.  Solid lines (---):  $\langle (\partial_x u_1)^3\rangle/\langle(\partial_x u_1)^2\rangle^{3/2}$.  Dashed lines ($--$): $\langle (\partial_y u_2)^3 \rangle/\langle(\partial_y u_2)^2\rangle^{3/2}$. Right: Kurtosis in the compression and expansion directions.  Solid lines (---):  $\langle (\partial_x u_1)^4\rangle/\langle(\partial_x u_1)^2\rangle^2$.  Dashed lines ($--$): $\langle (\partial_y u_2)^4 \rangle/\langle(\partial_y u_2)^2\rangle^2$.  In both plots: circles ($\bigcircle$), squares ($\medsquare$) and triangles ($\medtriangleup$): $S=1, 4, 10$ from the $R_{\lambda0} = 117$ set, and crosses ($\times$), asterisks (\textasteriskcentered) and pluses ($+$): $S=1, 4, 10$ from the $R_{\lambda0} = 193$ set.  An estimate of statistical error bar is shown and computed according to   (\ref{stderr}) with $X_j$ being the skewness  and kurtosis in the $j$-th realization. }
\label{skewandflat}
\end{figure}

The error bars in Figure~\ref{RstressAniso} indicate the statistical error of the quantities estimated from the finite number of realizations of the flow. That is, in $N$ realizations of the turbulent flow with a particular strain rate $S$, one obtains samples $\{ X_1, X_2, \dots, X_N\}$ of a quantity $X$.  The estimated standard error is the sample standard deviation of  $\{ X_1, X_2, \dots, X_N\}$ divided by $\sqrt{N}$.  That is, 
\begin{equation}
\frac{\sqrt{\sum_{j=1}^N (X_j-\overline{X})^2}}{\sqrt{N-1}\sqrt{N}}
\label{stderr}
\end{equation}
for data from $N$ realizations; here $\overline{X}$ is the mean of $\{ X_j\}_{j=1}^N$ .  In this work, the length of the symmetric error bars in Figures~\ref{RstressAniso} to \ref{ipaccvar2} is twice of the estimated statistical error.

Since tracer and inertial particle accelerations occur primarily at the smallest scales of motions, it is useful to look at the effects of the straining on the small scales. Figure \ref{du2dv2} shows a measure of the small scale anisotropy, the ratio of variances of longitudinal derivatives of the transverse and longitudinal velocity components $\langle (\partial_x u_2)^2\rangle/\langle(\partial_x u_1)^2\rangle$ with respect to time. At the highest strain rates, the anisotropy due to the straining is present at the smallest scales of motion, whereas for the lower strain rates, the flow appears nearly isotropic at the small scales. Note that the isotropic prediction for the ratio is $\langle (\partial_x u_2)^2\rangle/\langle(\partial_x u_1)^2\rangle=2$. The small scale anisotropy appears to become close to a constant after an initial transition period for the lower strain rates, but a stationary state is not reached at the highest rate of strain for this quantity.

The left hand side of Figure~\ref{skewandflat} shows the time evolution of longitudinal derivative skewness, along the directions of compression and expansion in the flow. Before the straining is applied, the skewness has a value of about $[-0.4,-0.5]$ as expected, but the straining causes a marked change in its value and becomes positive in the expanding direction. The effect in the compressed direction is more subtle, but an increase in magnitude (larger negative values) appear to occur for all strain rates given that the simulation is run for a sufficiently long time. The sign change indicates a change in the small scale structure of turbulence, namely that vortex structures dominate sheet-like structures resulting in an inhibition of the energy cascade (\cite{AW}). 

The right hand side of Figure~\ref{skewandflat} shows the longitudinal kurtosis, a measure of the flow intermittency. Here the effects are milder in both directions, although a small increase in the kurtosis is noted in the expanding directions (from the expected value of about 5-6, \cite{GAW}).  

\section{Particle acceleration statistics and discussion}

The non-uniform mean velocity field has a significant effect on the dynamics of tracers and inertial particles, both directly via the mean flow velocity and indirectly through the strained turbulent field.

\subsection{Tracers accelerations in straining flow}
The full acceleration of tracer particles in our flow is given by the equation: 
\begin{equation}
\widetilde{a}_{p_i} = \frac{D\tilde{u}_{p_i}}{Dt} = \frac{Du_{p_i}}{Dt} + u_{p_j} \frac{\partial U_{p_i}}{\partial x_j} + U_{p_j} \frac{\partial U_{p_i}}{\partial x_j}, \ \ i = 1,2,3.
\end{equation}
The subscript $p$ indicates that the full instantaneous flow velocity $\tilde{u}_{p_i}$, the fluctuating flow velocity $u_{p_j}$, and the mean flow velocity $ U_{p_i}$ are taken at the location of the tracer.
Here, we have assumed that the mean flow is time-independent. The mean of the tracer acceleration is $U_{p_j} \frac{\partial U_{p_i}}{\partial x_j}$; equal to the acceleration in a laminar flow of the same strain geometry. We are interested in the statistics of the tracer acceleration fluctuation, that is, when the pure mean flow contribution to the acceleration has been subtracted:
\begin{equation}
a_{p_i} = \frac{Du_{p_i}}{Dt} + u_{p_j} \frac{\partial U_{p_i}}{\partial x_j}, \ \ i = 1,2,3.
\label{tracerfluca}
\end{equation}
The first term on the right hand side refers to the material derivative of the fluctuating velocity field and represents the acceleration experienced by the fluid particle advected by the fluctuating flow field, and the second term refers to the tracer acceleration induced by turbulent transport in the mean velocity field.

In the variance of acceleration, the cross terms give rise to $\left\langle \frac{Du_{p_i}}{Dt}u_{p_i}  \right\rangle$ (no sum over $i$), representing the time derivative of the kinetic energy in the $i$-th component of velocity.  In addition, the latter term describes contributions of velocity variances, notably $(2S)^2\left\langle \left( u_{p_1}\right)^2 \right\rangle$ and $S^2\left\langle \left( u_{p_2}\right)^2 \right\rangle$ for the first and second component, respectively.  These terms are easily evaluated if the statistics of flow velocity fluctuations are available.

\subsubsection{Approximate tracer acceleration variances using rapid distortion theory}
When the straining is sufficiently rapid, the nonlinear and viscous forces vanish from the Navier-Stokes equations (e.g. see \cite{P_2000}), and therefore their solution is particularly convenient in comparison to solving the full Navier-Stokes equations. Below, we re-derive the RDT predictions for the evolution of the fluctuating velocity variances, as well as deriving the prediction of RDT on the evolution of the tracer acceleration variance.

In RDT, each Fourier mode evolves independently. Let us consider a single mode of the fluctuating velocity
\[
{\boldsymbol u}({\boldsymbol x}, t) = \hat{{\boldsymbol u}}_{\boldsymbol \kappa}(t) e^{i {\boldsymbol \kappa}(t) \cdot {\boldsymbol x}} \label{FSu}
\]
The wave number and the Fourier coefficients evolve according to the following set of equations (\cite{P_2000})
\begin{eqnarray}
\frac{d \kappa_\ell}{dt} &=& -\kappa_j\frac{\partial  U_j }{\partial x_\ell}, \label{kevo}\\
\frac{d \hat{u}_j}{dt} &=& - \hat{u}_k \frac{\partial  U_\ell }{\partial x_k}\left( \delta_{j\ell}-2\frac{\kappa_j\kappa_\ell}{|{\boldsymbol \kappa}|^2}\right). \label{ujevo}
\end{eqnarray} 
When the mean flow geometry, $ {\boldsymbol U}  = (-2Sx,Sy,Sz)$, has been applied, equation (\ref{kevo}), results in 
\begin{equation}
\kappa_\ell(t) = \kappa^0_\ell e^{-S_\ell t} \label{kappa},
\end{equation}
where $S_1 = -2S, S_2 = S_3 = S$, and $|{\boldsymbol \kappa}|^2 = (\kappa^0_1)^2 e^{4St} + (\kappa^0_2)^2 e^{-2St} +  (\kappa^0_3)^2 e^{-2St}$.
Similarly, for $\hat{u}_j$, equation (\ref{ujevo}), gives,
\[
 \frac{d \hat{u}_j}{dt} = - S_\ell\hat{u}_\ell \left( \delta_{j\ell}-2\frac{\kappa_j\kappa_\ell}{|{\boldsymbol \kappa}|^2}\right).
 \]
In particular, the long term prediction gives 
\[ 
\hat{u}_1(t) \approx \hat{u}_1^0 e^{-2St},\ \hat{u}_2(t) \approx \hat{u}_2^0 e^{-St}, \ \hat{u}_3(t) \approx \hat{u}_3^0 e^{-St}.
\]
Taking the material derivative of the single mode flow fluctuating velocity, we have
\begin{equation}
\frac{D u_1}{Dt} \approx -2S u_1, \ \frac{D u_2}{Dt} \approx -S u_2, \  \frac{D u_3}{Dt} \approx -S u_3.\label{RDTDuDt}	
\end{equation}
Therefore the anisotropic contribution to $\langle (Du_i/Dt)^2 \rangle$ can be estimated as $S_i^2 \langle (u_i)^2 \rangle$, where $S_1 = -2S, S_2 = S_3 = S$ are the strain rate at different directions.  Together with the normalized acceleration variances $a_0 \equiv (1/3)\langle a_ia_i\rangle / (\langle \epsilon_0 \rangle^3/\nu)^{1/2}$ at the beginning of the straining (\cite{VPCAB}), the magnitude of tracer fluctuating acceleration variance can be approximated as 
\begin{equation}
\langle (a_{p_i})^2 \rangle \approx a_0  \left(\frac{\langle \epsilon \rangle^3}{\nu}\right)^{1/2}+ 2S_i^2 \langle (u_i)^2 \rangle + S_i \frac{d \langle (u_i)^2 \rangle}{dt}, \ i = 1,2,3, \label{traceraccvarapprox}
\end{equation}
where the first term represents the isotropic contribution to acceleration variances with dependence on turbulence intensity, and all the terms on the right-hand side are Eulerian quantities of the straining flow.  Here $\langle \epsilon_0 \rangle$ is the mean energy dissipation rate in homogeneous isotropic turbulence across all realizations, and $\langle \epsilon \rangle$ is the time-dependent mean energy dissipation rate in the straining turbulence across all realizations. 

\begin{figure}
\vspace*{3mm}
\hspace*{-2mm}
\begin{center}
\includegraphics[width=2.66in]{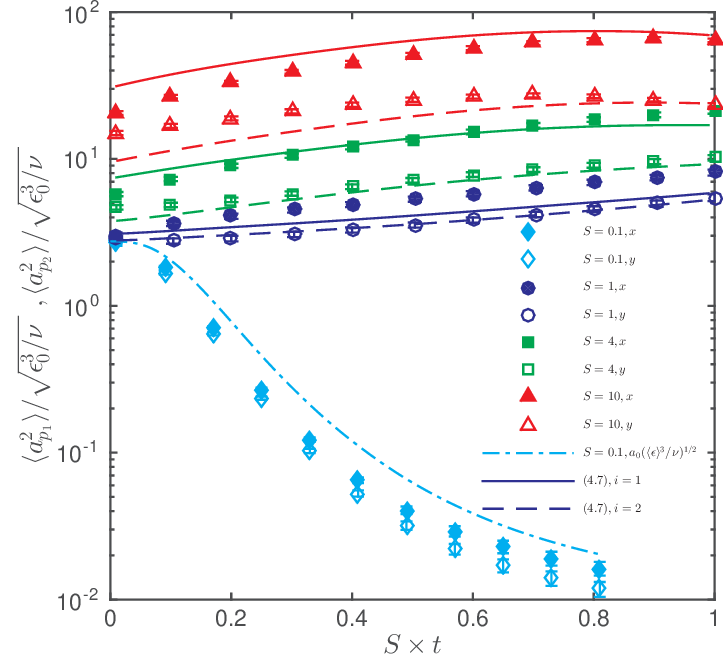}
\includegraphics[width=2.6in]{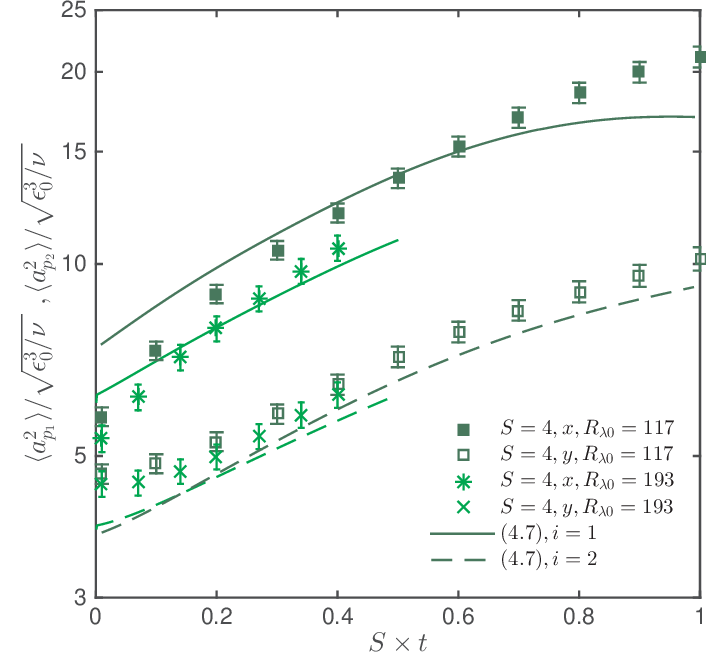}
\caption{Normalized acceleration variances of passive tracers in the compressed and expanding directions vs. strain time at various strain rates.  Solid and empty symbols represent the compressed and expanding directions, respectively.  Solid (---) and dashed ($--$) lines indicate RDT long term predictions of the tracer acceleration variances from the straining together with normalized flow acceleration variances in HIT shown in (\ref{traceraccvarapprox}).  Dash-dot line ($-.$) indicates the term $ a_0  /\left({\langle \epsilon \rangle^3}/{\nu}\right)^{1/2}$ for the $S=0.1$ case.  Diamond ($\meddiamond$), circle ($\bigcircle$), square ($\medsquare$), triangle ($\medtriangleup$) mark normalized tracer acceleration variances in strain rates $S = 0.1, 1, 4$ and $10$.  Left: tracer acceleration variances in different strain rates in $R_{\lambda0} = 117$.  Right: tracer acceleration variances in $S=4$ in $R_{\lambda0} = 117$ (squares ($\filledmedsquare,\medsquare$): compressed and expanding directions, respectively) and $193$ (asterisks (\textasteriskcentered)  and pluses ($+$): compressed and expanding directions, respectively).  An estimate of statistical error bar is shown and computed according to   (\ref{stderr}) with $X_j$ being $\langle a_{p_i}^2 \rangle/\sqrt{\epsilon_0^3/\nu}$ in the $j$-th realization. }
\label{traceraccvar}
\end{center}
\end{figure}

Figure \ref{traceraccvar} shows the acceleration variances of passive tracers in the straining flow. The variance is higher in the compression direction than in the expanding directions due to the mean straining geometry, and the effects are seen immediately after the strain has been imposed. The approximations derived above, applying RDT, fit nicely to the simulation data as shown. The change in the isotropic dissipation rate due to straining is accounted for in the first term of (\ref{traceraccvarapprox}), and the rest of the terms involve the mean strain and the velocity fluctuations. The mean strain causes a rapid increase in the acceleration variance as the strain is applied, particularly at the higher rate of strain. The subsequent increase and the differences between the individual components are partially due to the evolution of the velocity variances for each component; namely that the compressed velocity components are emphasized (increasing energy content) whereas the expanding velocity components are either suppressed or maintained.  

\subsection{Inertial particle accelerations in straining flow}
The differential equations for the particle position as a function of time, obtained by combining the mean flow components in equations (\ref{meanflow}) with the equations (\ref{particleloc}) and (\ref{particlevel}) are second order linear ordinary differential equations with constant coefficients:
\begin{eqnarray*}
&& \frac{d^2 x_{p_i}}{dt^2} +\frac1{\tau_p} \frac{dx_{p_i}}{dt}- \frac{S_i}{\tau_p}x_{p_i} = \frac1{\tau_p}u_i,\  \ i= 1,2,3.\\
\end{eqnarray*}
The roots $\displaystyle \lambda_{1,2} = \frac{-1\pm\sqrt{1+4S_i \tau_p}}{2\tau_p}$ of the characteristic equations of these equations prescribe the behavior of particle movements in absence of turbulence. In the compression direction, the combination of strain rates $S$ and the Stokes relaxation time $\tau_p$ in our simulations give rise to the discriminant $D_1=1-8S\tau_p$, which separates two possible movements: a overdamped-decay motion toward the stagnation plane $x=0$ or a underdamped-oscillation about the stagnation plane.  In the expanding direction the discriminant $D_{2,3} = 1+4S\tau_p$ is always positive, and particles move away from the axis $y=z=0$ exponentially in time when only mean flow is considered.  The accelerations of particles follow a similar patten.  

When turbulent fluctuations are present in the strained flow, the statistical description of the dynamics of the inertial particles becomes much more complicated. Treating the fluctuating flow velocity as a source term, one can solve the equations formally using the Laplace transform.  Specifically, the formal expression of particle accelerations are:\\
In the compression direction:
\begin{eqnarray}
D_1 > 0: && \nonumber \\
\widetilde{a}_{p_1}(t) &=& \frac1{\lambda_1-\lambda_2}\left[(\lambda_2^2(\lambda_1+2S)e^{\lambda_2 t} -\lambda_1^2(\lambda_2+2S)e^{\lambda_1 t})x_{p0} + (\lambda_1^2e^{\lambda_1t} -\lambda_2^2 e^{\lambda_2t}){v}_{p_10} \right.\nonumber\\
&&+ \left.  \frac1{\tau_p}\int_0^t u_1({\boldsymbol x}_p(\tau),\tau)(\lambda_1^2e^{\lambda_1(t-\tau)}-\lambda_2^2e^{\lambda_2 (t-\tau)})\ d\tau \right] +\frac{1}{\tau_p}u_1({\boldsymbol x}_p(t),t).\label{a1fullDp}
\end{eqnarray}
\begin{eqnarray}
D_1 < 0: && \nonumber \\
\widetilde{a}_{p_1}(t) &=& e^{-\frac{t}{2\tau_p}}\left[-\frac1{\tau_p}{v}_{p_10}\cos(\omega t) + \frac1{2\tau_p^2 \omega}((1-4S\tau_p){v}_{p_10}+8S^2\tau_px_{p0})\sin(\omega t)\right.\nonumber\\
&&-\left.\frac1{\tau_p^2}\int_0^t u_1({\boldsymbol x}_p(\tau),\tau)e^{\frac{\tau}{2\tau_p}}\left(\cos(\omega (t-\tau))+\frac{4S\tau_p-1}{2\tau_p\omega}\sin(\omega (t-\tau))\right)\right]\nonumber\\
&&+\frac1{\tau_p}u_1({\boldsymbol x}_p(t),t).\label{a1fullDn}
\end{eqnarray}
In the expressions, 
\begin{equation}
\lambda_{1,2}  = -\frac1{2\tau_p} (1\pm\sqrt{1-8S\tau_p}),\ \ \omega=\frac1{2\tau_p}\sqrt{8S\tau_p-1} \label{xlambdas}
\end{equation}
as well as the Stokes relaxation time $\tau_p$ determine time scales for various contributions to inertial particle accelerations. $x_{p0}$ and $v_{p_10}$ are position and velocity of the particle right before the straining starts (at $t=0^-$). In the expanding direction $y$:  (the expression in the $z$ direction is similar)
\begin{eqnarray}
\widetilde{a}_{p_2}(t) &=& \frac1{\lambda_{1}-\lambda_{2}}\left[ (\lambda_{2}^2(\lambda_1-S)e^{\lambda_{2} t} -\lambda_{1}^2(\lambda_2-S)e^{\lambda_{1} t})y_{p0} + (\lambda_{1}^2e^{\lambda_{1}t} -\lambda_{2}^2 e^{\lambda_{2}t}){v}_{p_20} \right.\nonumber\\
&&+ \left.  \frac1{\tau_p}\int_0^t u_2({\boldsymbol x}_p(\tau),\tau)(\lambda_{1}^2e^{\lambda_{1} (t-\tau)}-\lambda_{2}^2e^{\lambda_{2} (t-\tau)})\ d\tau \right] +\frac{1}{\tau_p}u_2({\boldsymbol x}_p(t),t). \label{a2full}\end{eqnarray}
Similarly, $y_{p0}$ and $v_{p_20}$ are position and velocity of the particle at $t=0^-$, and
\begin{equation}
\lambda_{1,2}  = -\frac1{2\tau_p} (1\pm\sqrt{1+4S\tau_p}). \label{ylambdas}
\end{equation}

The mean flow influences the variances when full acceleration is considered, and so is the case for other statistics involving full particle velocity or accelerations. Since the magnitude of the mean flow velocity depends on the location in the domain, the magnitudes in the variances of acceleration components are characterized by the domain size in respective directions (through the initial positions $x_{p0}$, $y_{p0}$ in acceleration expressions (\ref{a1fullDp}), (\ref{a1fullDn}) and (\ref{a2full})) in addition to the rate of strain. 

In an attempt to minimize the influence of the mean flow, in addition to ensuring that our statistics are deduced from sufficient many independent samples, we condition our inertial particle analysis on particles that started in a thin layer parallel to and next to the $x=0$ plane for the $x$-component statistics and a thin layer parallel to and next to the $y=0$ plane for the $y$-component statistics. For the strain rates $S =0.1$ and $S=1$, we use layers with thickness of 8 lattice units in the $R_{\lambda0} = 117$ simulations.  For $S=4$ and $S=10$ flows, we use layers with thicknesses of 4 and 2 lattice units.  With the selected thicknesses, we ensure the difference of the mean flow velocities within the layers do not exceed 55\% of  $u_1^{rms}$ in the compression direction, and do not exceed 77\% of $u_2^{rms}$ in the expanding direction.  The numbers of particles available in the layers decreases with the thickness of the layers from about 61,500 in $S=1$ flows to 15,500 in $S=10$ flows in the compression direction.  In the expanding direction, the number of particles used for the analysis decreases from 240,000 in $S=1$ to 60,700 in $S=10$.  We also apply the symmetry with respect to the planes $x=0$ and $y=0$.

\begin{figure}
\vspace*{5mm}
\begin{center}
\includegraphics[width=5.4in,height=2.5in]{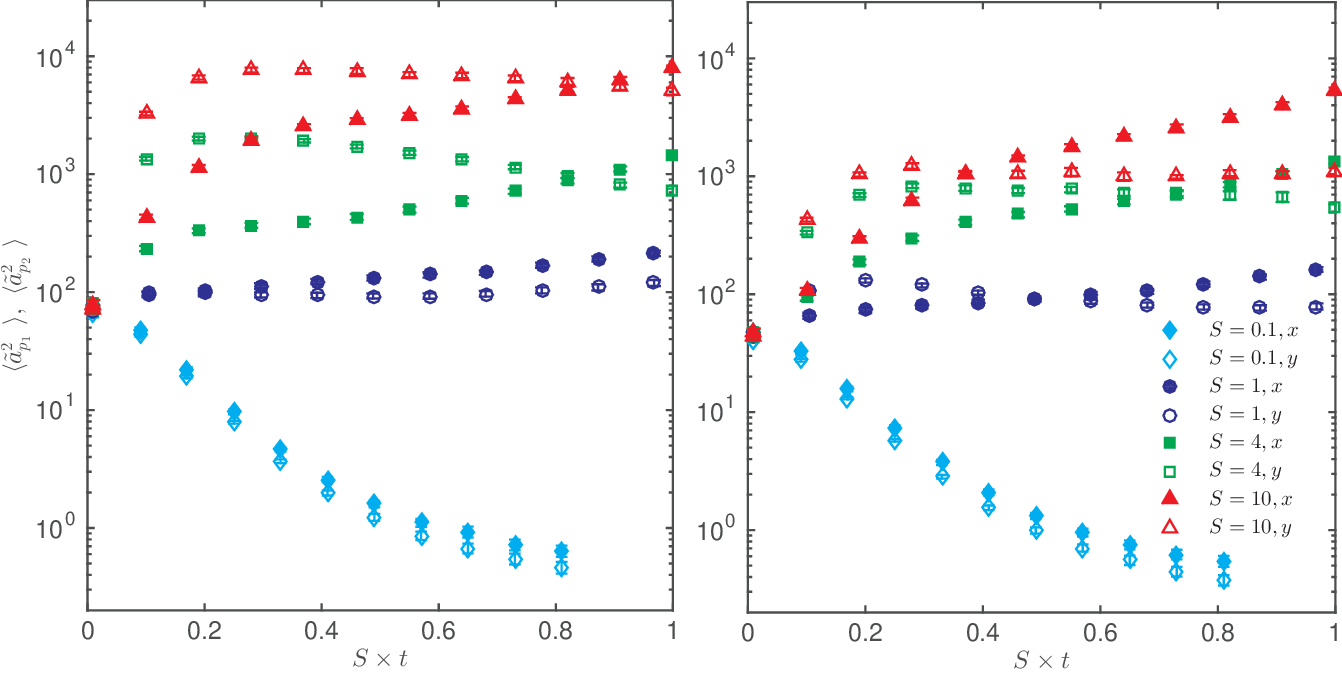}
\caption{Acceleration variances of inertial particles in the compressed (solid symbols) and expanding (empty symbols) directions vs. strain time for two types of particles Left: $\tau_p = 0.015$; Right and $\tau_p=0.05$. Symbols refer to different strain rates, diamond ($\meddiamond$) $S=0.1$; circle ($\bigcircle$) $S=1$; square ($\medsquare$) $S=4$;  triangle ($\medtriangleup$) $S=10$. The data are from the data set $R_{\lambda0} = 117$.  An estimate of statistical error bar is shown and computed according to   (\ref{stderr}) with $X_j$ being particle acceleration variances in the $j$-th realization.}
\label{ipaccvar}
\end{center}
\end{figure}

Figure \ref{ipaccvar} shows the acceleration variances of inertial particles in the straining flow for two particle types $\tau_p=0.015$ and $\tau_p=0.05$, which correspond to $St = 0.3$ and $1$ at the beginning of the straining.  These two types of particles have positive discriminants in $S=1$, correspond to $D_1 >0$ and $D_1 < 0$ in $S=4$, and have negative discriminants in  $S=10$. 

\subsubsection{Initial transition period of acceleration variances}
\begin{figure}
\begin{center}
\vspace*{2mm}
\includegraphics[height=1.91in]{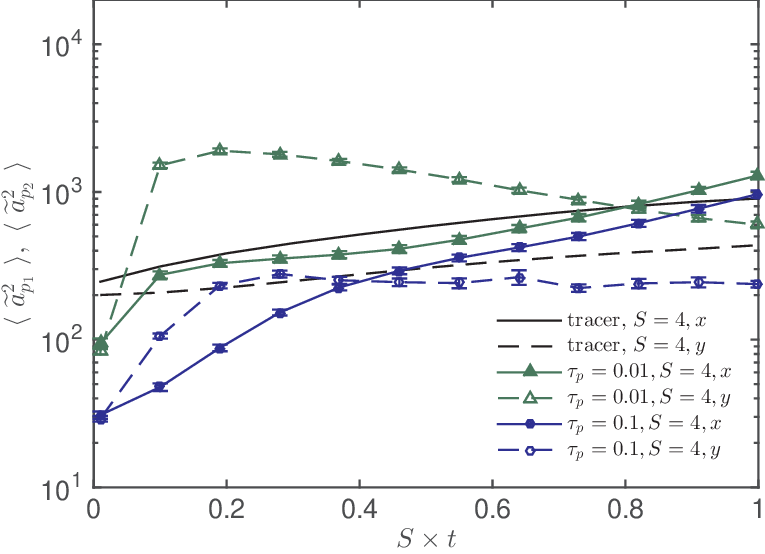}
\hspace*{2.2mm}
\includegraphics[height=1.91in]{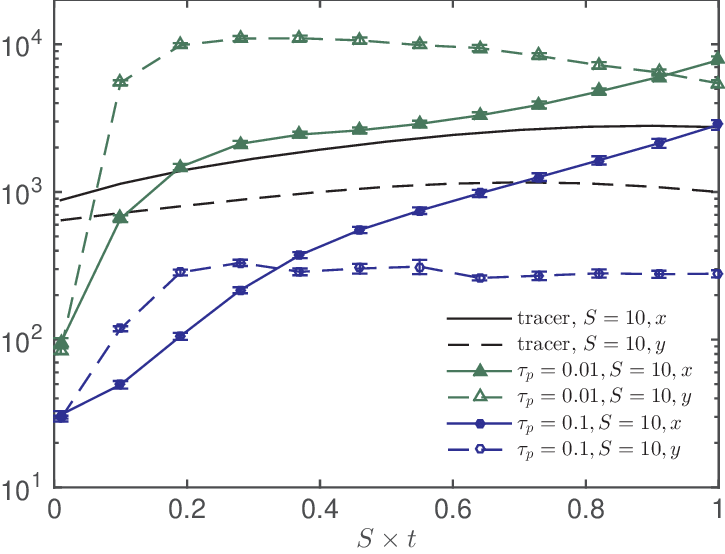}
\end{center}
\caption{Acceleration variances for tracers and inertial particles with $\tau_p = 0.01, 0.1$.  Lines with circles ($\bigcircle$): $\tau_p = 0.1$; lines with triangles ($\medtriangleup$): $\tau_p = 0.01$.  Solid lines with solid symbols: $\langle (\widetilde{a}_{p_1})^2 \rangle$; dashed lines with empty symbols: $\langle (\widetilde{a}_{p_2})^2 \rangle$.  Solid lines (---): tracers $\langle ({a}_{p_1})^2 \rangle$; dashed lines ($--$): tracers $\langle ({a}_{p_2})^2 \rangle$. Left: $S=4$, Right: $S=10$. The data are from the data set $R_{\lambda0} = 117$.  An estimate of statistical error bar is shown and computed according to   (\ref{stderr}) with $X_j$ being particle acceleration variances in the $j$-th realization.}
\label{ipaccvar2}
\end{figure}

The acceleration variances of the inertial particles show a transition period at the beginning of the straining that is not seen in the tracer accelerations.  Equations (\ref{a1fullDp}) and (\ref{a1fullDn}) indicate that the initial position ${x}_{p0}$ and velocity ${ v}_{p_10}$ of an inertial particle affects its acceleration.  The time scale of this influence depends on the  exponents of the exponential terms in the acceleration expressions.  In the compression direction, if $D_1 < 0$ the decaying rate is $2\tau_p$.  If $D_1 >0$, both $\lambda_1$ and $\lambda_2$ are negative so the decaying rate is $\min(\frac1{|\lambda_1|},\frac1{|\lambda_2|}) = \frac1{|\lambda_2|}$.  This explains the different lengths of the initial transition period for particles with different Stokes numbers in flows with a fixed strain rate.  In the left figure of Figure \ref{ipaccvar2}, we show acceleration variances of particles with $\tau_p = 0.01$ and $0.1$ in $S=4$ to elucidate the time scale for initial transition.  The ratio between the decaying rates of these two types of particles is about 1.75. 

Although $\langle (u_2)^2 \rangle$ decreases with straining, $\langle \widetilde{a}_{p2}^2 \rangle$ increases at the beginning and a maximum occurs. This is because $\lambda_2$ is positive in the expanding direction. Similar to the case discussed in the $x$-direction, the exponents determine the time scale of the behavior of acceleration variances.   From (\ref{a2full}) one can estimate the time that the maximum takes place by ignoring the flow fluctuation, and obtain the ratio between the times of maximum acceleration variances of the two particles presented in the figure to be 1.52.  It appears that our simulation data confirms such an estimation.  For $S=10$, our simulations were too short to display the post-transition period.

\subsubsection{Magnitude of the acceleration variances}

Figure \ref{ipaccvar} shows that the acceleration variances of inertial particles increases with the strain rate.  This is mainly due to the mean flow contributing to the acceleration in terms of the factors $2S$ and $S$ in the expressions (\ref{a1fullDp}), (\ref{a1fullDn}) and (\ref{a2full}) and, in a smaller part, through the exponents that regulate the influence of the initial conditions.  

At the onset of straining, the acceleration variances in the expanding direction is higher than that in the compression direction.  This could be explained through the exponents of the kernels.  In the compression direction, the coefficients $\lambda_1, \lambda_2$ and $-\frac1{2\tau_p}$ in the exponents of the kernels in (\ref{a1fullDp}) and (\ref{a1fullDn}) are all negative and indicate decay with time.  But for the expanding direction, $\lambda_2 = \frac{-1+\sqrt{1+4S\tau_p}}{2\tau_p}$ is positive, which leads to the increase of magnitude.  For sufficiently long time, the influence of the initial condition is diminished.  The magnitude of the flow velocity fluctuation (in the last two terms of the acceleration expressions) in the expanding direction remains roughly constant, while in the compression direction the flow velocity fluctuation grows exponentially. As a result the acceleration variances in the compression direction overtakes that in the expanding direction.

In contrast to the isotropic homogeneous turbulence (\cite{AGSWCB, BBBCCLMT}), the acceleration variances of inertial particles do not necessary have lower magnitudes compared with that of tracers in strained flow.  The inertial particle with $\tau_p= 0.01$ in the right plot of Figure \ref{ipaccvar2} depicts such a situation.  Recall that (\ref{traceraccvarapprox}) provides an approximation of tracer acceleration variances, and in higher strain rates the terms $2S_i^2 \langle (u_i)^2 \rangle + S_i d\langle (u_i)^2 \rangle/dt$ are the main contributor of the variances.  From Figure \ref{RstressAniso}  we know that the variance of flow fluctuating velocity grows exponentially in time, so its time derivative can be estimated as a constant multiple $c_1S$ ($c_1$ is a constant) of the variance itself.   Hence the main terms in the tracer acceleration variance increases in the order of $S_i^2\langle (u_i)^2 \rangle$ with the strain rate.  For the inertial particles, however, their long term variances depends on the term $\langle (u_i)^2 \rangle/\tau_p^2$.  When $\tau_p$ is small enough, the factor $1/\tau_p^2$ is larger than the magnitude of $S_i^2$, and the inertial particle acceleration variances surpass the tracer acceleration variances.

\subsection{Probability density functions of particle accelerations}

\begin{figure}
\vspace*{8mm}
\begin{center}
\includegraphics[width=5.4in]{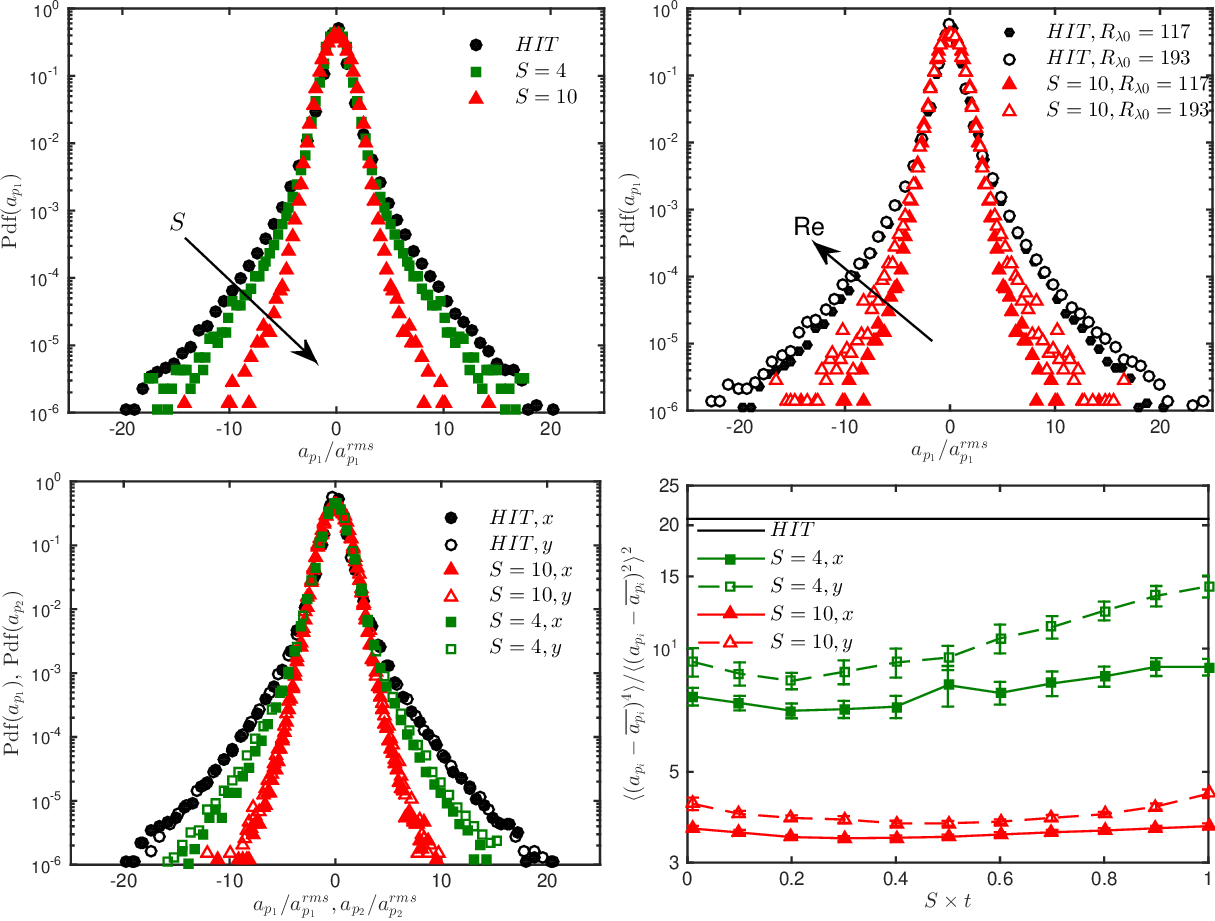}
\end{center}
\caption{Normalized pdfs of tracer accelerations and tracer acceleration flatness.  Top left: Pdf($a_{p_1}$) in HIT (circle \textbullet), straining flow $S=4$ (square $\filledmedsquare$), and $S=10$ (triangle $\filledmedtriangleup$).  Top right: Pdf($a_{p_1}$) in HIT (circle \textbullet: $R_{\lambda0} =117$, $\circ: R_{\lambda0} = 193$) and in straining flow $S=10$ (trianlge $\filledmedtriangleup: R_{\lambda0} =117$, $\medtriangleup: R_{\lambda0} = 193$).  Bottom left: Pdf($a_{p_1}$) and Pdf($a_{p_2}$) in $R_{\lambda0}=117$ flow.  Circles (\textbullet, $\circ$): HIT, squares ($\filledmedsquare, \medsquare$): $S=4$, triangles ($\filledmedtriangleup, \medtriangleup$): $S=10$.  Solid symbols: $i=1$ component, empty symbols: $i=2$ component. Bottom right: Tracer acceleration flatness,$\langle\left(a_{p_i}-\overline{a_{p_i}} \right)^4\rangle/\langle\left(a_{p_i}-\overline{a_{p_i}} \right)^2\rangle^2$ at $R_{\lambda0} =117$.  Solid line (---): HIT,  squares ($\filledmedsquare, \medsquare$): $S=4$, triangles ($\filledmedtriangleup, \medtriangleup$): $S=10$.  Solid symbols: $i=1$ component, empty symbols: $i=2$ component.}
\label{traceraccpdf}
\end{figure}

\begin{figure}
\vspace*{5mm}
\begin{center}
\includegraphics[height=4.2in,width=5.4in]{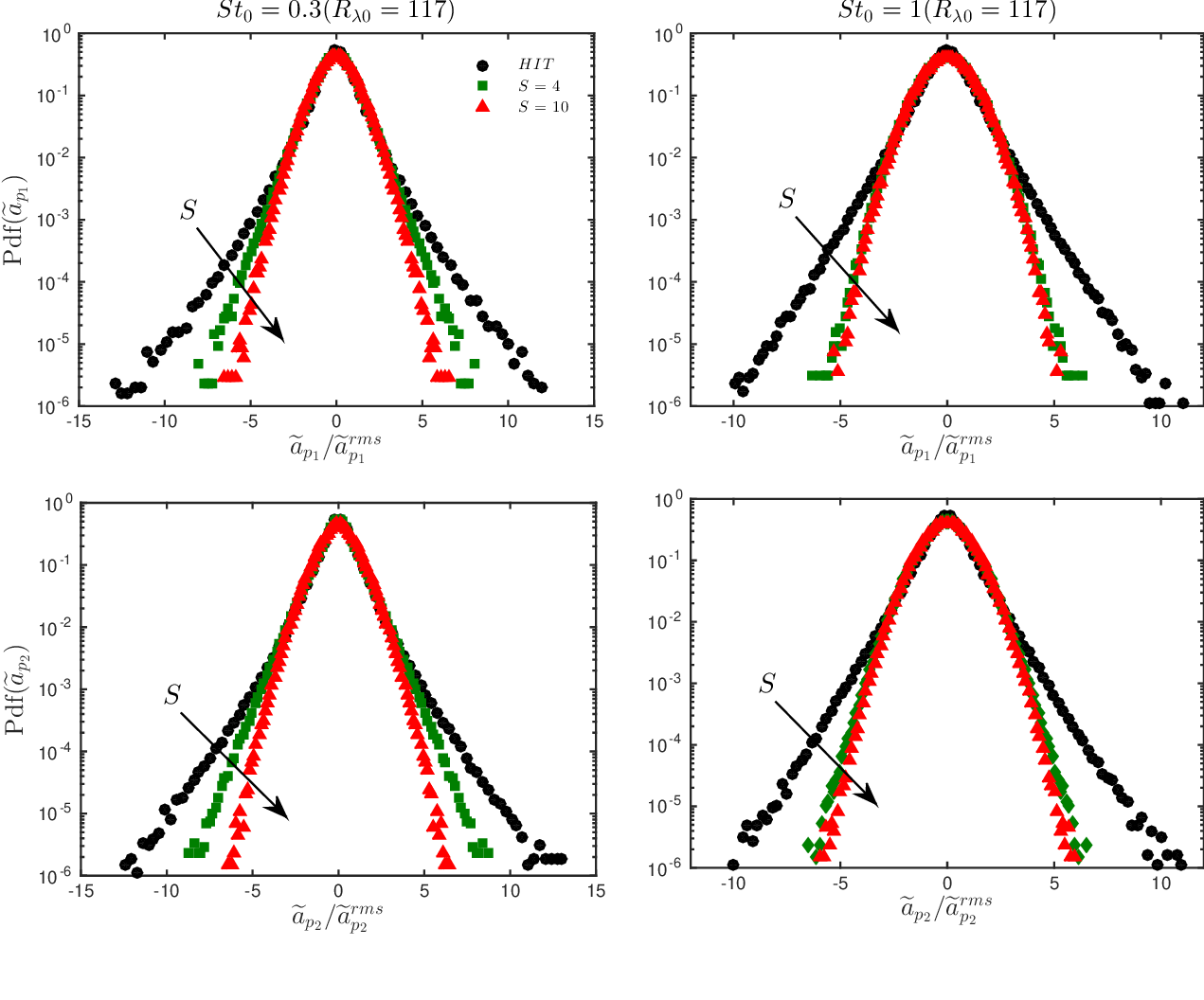}
\vspace*{-8mm}
\caption{Normalized probability distribution function of particle acceleration components $\widetilde{a}_{p_i}$ for different strain rates, for particles originating in thin slices parallel to the $x=0$ plane, for the $i=1$ component and to the $y=0$ plane for the $i=2$ component. Left:  particles with $St_0=0.3$ in $R_{\lambda0}=117$. Right: particles with $St_0=1$ in $R_{\lambda0}=117$. Top: The $i=1$ component. Bottom: The $i=2$ component. Circles (\textbullet): HIT; squares ($\filledmedsquare$): $S=4$; triangles ($\filledmedtriangleup$): $S=10$.}
\label{iplyaccpdf_S}
\end{center}
\end{figure}

\begin{figure}
\vspace*{5mm}
\begin{center}
\includegraphics[height=4.2in,width=5.4in]{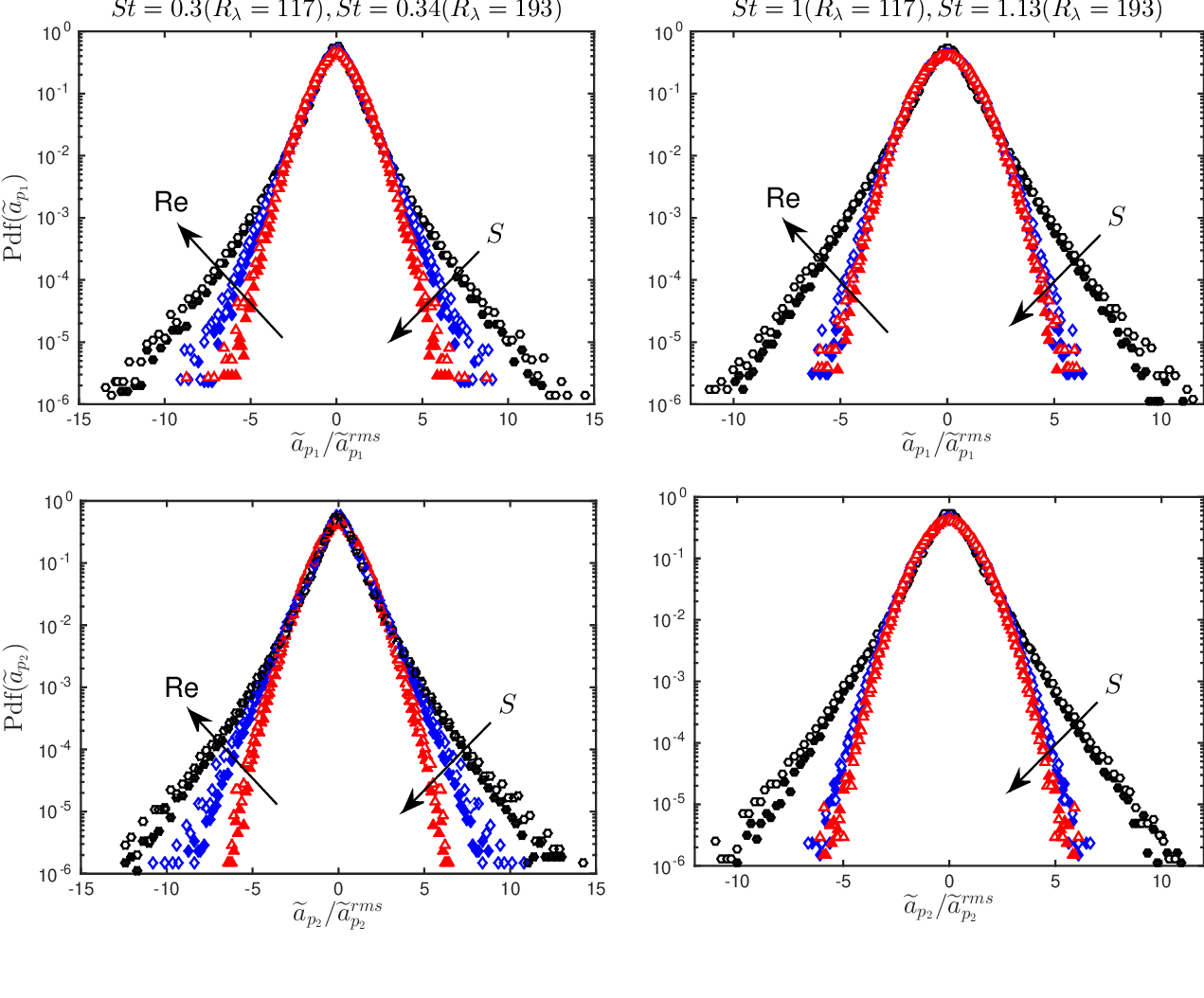}
\caption{Normalized probability distribution function of particle acceleration components $\widetilde{a}_{p_i}$ for two Reynolds numbers, for particles originating in thin slices parallel to the $x=0$ plane, for the $i=1$ component and to the $y=0$ plane for the $i=2$ component. Left:  particles with $St_0=0.3$ in $R_{\lambda0}=117$ and $St_0=0.34$ in $R_{\lambda0}=193$. Right: particles with $St_0=1$ in $R_{\lambda0}=117$ and $St_0=1.12$ in $R_{\lambda0}=193$. Top: The $i=1$ component. Bottom: The $i=2$ component. Circles (\textbullet, $\circ$): HIT; triangles ($\filledmedtriangleup,\medtriangleup$): $S=10$.  Solid symbols: $R_{\lambda0}=117$; empty symbols: $R_{\lambda0}=193$.}
\label{iplyaccpdf_Re}
\end{center}
\end{figure}

Figures \ref{traceraccpdf} to \ref{iplyaccpdf_Re} show the particle acceleration pdfs at various rates of strain ($S=1,4$ and $10$) and at two Reynolds numbers ($R_{\lambda0}=117$ and $R_{\lambda0}=193$) for tracers and inertial particle ($\tau_p=0.015$ and $\tau_p=0.05$), respectively. The effect of increasing the rate of strain is most notably seen by the narrowed pdf-tails for the tracer and inertial particle accelerations. 

The top left plot of Figure \ref{traceraccpdf} demonstrates the narrowing of the tracer acceleration pdfs in straining flow.  The top right plot shows that the narrowing effect is milder at the higher Reynolds number due to the faster time response of the smaller scales at the higher Reynolds number. The increase in the magnitude of acceleration variances due to the mean straining appears to be the primary reason for the tail-narrowing. The bottom left plot of Figure \ref{traceraccpdf} indicates the response of acceleration in the compressed and expanding directions. Although the flow field is very different component wise, the acceleration pdfs, resulting from tracers following trajectories of small scale structures in the fluid are more or less identical. Additionally, the evolution of the acceleration flatness $\langle\left(a_{p_i}-\overline{a_{p_i}} \right)^4\rangle/\langle\left(a_{p_i}-\overline{a_{p_i}} \right)^2\rangle^2$ (a measure of the intermittency of the acceleration) is shown in the bottom right figure for the lower Reynolds number $R_{\lambda0} =117$, emphasizing the narrowing effect of straining and the difference between components. 

In terms of inertial particles, as for isotropic homogeneous turbulence, the narrowing of the pdf tails follows an increase in the Stokes number.  Figure \ref{iplyaccpdf_S} illustrates the narrowing effects. Such behavior has been demonstrated in a number of previous studies(\cite{AGSWCB, BBBCCLMT}), owing to the heavier particles passing through, or selectively filtering, the most rapid motions in the flow field. However, here the situation is more complex, since the acceleration variance is not necessarily smaller for inertial particles, due to the strong effect of the mean flow as discussed above.  Figure \ref{iplyaccpdf_Re} shows that the higher Reynolds number has a expanding effect on lighter inertial particle acceleration pdf tails.  However, for the heavier particles ($St=1$) in higher strain $S=10$, the effect is not as evident in the expanding direction. We note that the Stokes number of a particle increases slightly during the straining due to a decrease in the Kolmogorov timescale (less than $ 10$\% for the highest rate of strain). This effect contributes to the narrowing of the acceleration pdf tails, but to a lesser effect than the increased variances.

It is interesting to consider our results in a wider context of flows with non-zero mean components, for example by comparing with the dynamics in shear flow. \cite{LSGWC2010} and \cite{GSNW_2008} considered tracers and inertial particles in a non-uniform shear, namely a turbulent boundary layer. An increased rate of shear, closer to the wall, resulted in an increased acceleration variance in the lateral component but a milder effect in the transverse component, which appears to be consistent with the predictions of equation (\ref{tracerfluca}) for their geometry. As a result of the increased acceleration variances, attenuation was found in the tails of the inertial particle acceleration pdfs as observed in this work. 

\section{Conclusions}
Our results show a strong effect of the mean flow straining on the Lagrangian acceleration statistics, both for the passive tracers and the inertial particles. The effect of straining is primarily felt in acceleration variances and pdfs when the rate of strain is sufficiently high such that the strain timescale is of a comparable magnitude to the Kolmogorov timescale. For high rates of strain the magnitude of the acceleration variances are increased significantly and the tails of the normalized acceleration pdfs for tracers and inertial particles narrow. The former effect is well explained by observing the predicted behavior of the acceleration variance by rapid distortion theory. RDT provides us a relation between the flow velocity variances and its acceleration variances and illustrates the dependence of the acceleration variance on the rate of strain. 

However, the effect is complex, partly due to the connection, or lack thereof, between particle acceleration component in one direction and the fluid flow in the same direction; a particle trajectory around a strong vortex will result in large acceleration values in the directions normal to the axis of the vortex. But there is also a direct contribution from the mean straining and fluctuating velocity in the acceleration, resulting for example in an increased variance value of acceleration in both compressing and expanding directions.

For tracers, the narrowing of the normalized acceleration pdfs stems therefore in a complex manner from both of these effects.  The same effect is also felt by the lighter inertial particles, given that their inertia is sufficiently small to sample the small scale motions. Because of their small inertia the interplay with the mean flow enable their acceleration variances to rise beyond that of tracers.  

When the inertia is further increased the ballistic particle motion in the rapidly accelerating mean flow becomes increasingly important leading to Gaussian pdf tails. Here, the particles are swept through the fluid, and the slower, large scales of motions are more likely to influence the particles. This could also be seen by the lower magnitude of acceleration variances of the heavier particles compared with the lighter inertial particles. We derive the formal expressions for inertial particle acceleration, and these expressions reveal the complex interplay between flow straining and particle inertia. 

Our findings emphasize the importance of the presence of strong mean motions and imposed small scale anisotropy in particle laden flows. It is our opinion that the results have relevance to the understanding and modelling of a range of practical deforming or straining flows where inertial particles are important aspects of the process.  In particular, we believe that our findings may help in the development of sub-grid Lagrangian models for particles in the proximity of straining regime near stagnation points.

We thank Zellman Warhaft, Paolo Gualtieri, and Michel van Hinsberg for their comments and suggestions. C. Lee acknowledges the hospitality of the School of Science and Engineering at Reykjav\'ik University and the Department of Mathematics and Computer Science and the Department of Applied Physics at Eindhoven University of Technology. This work was supported by the Icelandic Research Fund and  was partially supported by a research program of the Foundation for Fundamental Research on Matter (FOM), which is part of the Netherlands Organisation for Scientific Research (NWO).  Support from COST Actions MP0806 and MP1305 is also kindly acknowledged.


\begin{thebibliography}{29}
\expandafter\ifx\csname natexlab\endcsname\relax\def\natexlab#1{#1}\fi

\bibitem[Ayyalasomayajula {\em et~al.\/}(2006)Ayyalasomayajula, Gylfason,
  Salazar, Warhaft, Collins \& Bodenschatz]{AGSWCB}
{\sc Ayyalasomayajula, S., Gylfason, A., Salazar, J., Warhaft, Z., Collins,
  L.~R. \& Bodenschatz, E.} 2006 Inertial particle accelerations in turbulence:
  Experiments and numerical simulations. In {\em ASME Fluids Conference\/}.

\bibitem[Ayyalasomayajula \& Warhaft(2006)]{AW}
{\sc Ayyalasomayajula, S. \& Warhaft, Z.} 2006 Nonlinear interactions in
  strained axi-symmetric high \mbox{R}eynolds number turbulence. {\em J. Fluid
  Mech.\/} {\bf 566}, 273--307.

\bibitem[Batchelor(1953)]{B_1953}
{\sc Batchelor, G.K} 1953 {\em The Theory of Homogeneous Turbulence\/}.
  Cambridge University Press.

\bibitem[Bec {\em et~al.\/}(2006)Bec, Biferale, Boffetta, Celani, Cencini,
  Lanotte, Musacchio \& Toschi]{BBBCCLMT}
{\sc Bec, J., Biferale, L., Boffetta, G., Celani, A., Cencini, M., Lanotte, A.,
  Musacchio, S. \& Toschi, F.} 2006 Acceleration statistics of heavy particles
  in turbulence. {\em J. Fluid. Mech.\/} {\bf 550}, 349.

\bibitem[Biferale \& Procaccia(2005)]{BP_2005}
{\sc Biferale, L. \& Procaccia, I.} 2005 Anisotropy in turbulent flows and in
  turbulent transport. {\em Physics Reports\/} {\bf 414}~(2-3), 43 -- 164.

\bibitem[Celani(2007)]{C_2007}
{\sc Celani, A.} 2007 The frontiers of computing in turbulence: challenges and
  perspectives. {\em J. Turbul.\/} {\bf 8}, N34.

\bibitem[Chen {\em et~al.\/}(2006)Chen, Meneveau \& Katz]{CMK}
{\sc Chen, J., Meneveau, C. \& Katz, J.} 2006 Scale interaction of turbulence
  subjected to a straining-relaxation-destraining cycle. {\em J. Fluid Mech.\/}
  {\bf 562}, 123--150.

\bibitem[Gerashchenco {\em et~al.\/}(2008)Gerashchenco, Sharp, Neuscamman \&
  Warhaft]{GSNW_2008}
{\sc Gerashchenco, S., Sharp, N.~S., Neuscamman, S. \& Warhaft, Z.} 2008
  Lagrangian measurements of inertial particle accelerations in a turbulent
  boundary layer. {\em J. Fluid Mech.\/} {\bf 617}, 255--281.

\bibitem[Gualtieri \& Meneveau(2010)]{GM_2010}
{\sc Gualtieri, P. \& Meneveau, C.} 2010 Direct numerical simulations of
  turbulence subjected to a straining and destraining cycle. {\em Phys.
  Fluids\/} {\bf 22}, 065104.

\bibitem[Gualtieri {\em et~al.\/}(2009)Gualtieri, Picano \&
  Caisciola]{GPC_2009}
{\sc Gualtieri, P., Picano, F. \& Caisciola, C.M.} 2009 Anisotropic clustering
  of inertial particles in homogeneous shear flow. {\em J. Fluid Mech.\/} {\bf
  629}, 25--39.

\bibitem[Gualtieri {\em et~al.\/}(2012)Gualtieri, Picano, Sardina \&
  Casciola]{GPSC_2012}
{\sc Gualtieri, P., Picano, F., Sardina, G. \& Casciola, C.M.} 2012 Statistics
  of particle pair relative velocity in the homogeneous shear flow. {\em
  Physica D\/} {\bf 241}, 245--250.

\bibitem[Gylfason {\em et~al.\/}(2004)Gylfason, Ayyalasomayajula \&
  Warhaft]{GAW}
{\sc Gylfason, A., Ayyalasomayajula, S. \& Warhaft, Z.} 2004 Intermittency,
  pressure and acceleration statistics from hot-wire measurements in
  wind-tunnel turbulence. {\em J. Fluid Mech.\/} {\bf 501}, 213--229.

\bibitem[Gylfason {\em et~al.\/}(2011)Gylfason, Lee, Perlekar \&
  Toschi]{GLPT_2011}
{\sc Gylfason, A., Lee, C., Perlekar, P. \& Toschi, F.} 2011 Direct numerical
  simulation on strained turbulent flows and particles within. {\em Journal of
  Physics: Conference Series\/} {\bf 318}, 052003.

\bibitem[Han \& Reitz(1995)]{HR_1995}
{\sc Han, Z. \& Reitz, R.D.} 1995 Turbulence modeling of internal combustion
  engines using rng $\kappa-\epsilon$ models. {\em Combustion Science and
  Technology\/} {\bf 106}, 267--295.

\bibitem[Hunt(1973)]{H1973}
{\sc Hunt, J.C.R.} 1973 A theory of turbulent flow round two-dimensional bluff
  bodies. {\em J. Fluid. Mech.\/} {\bf 61}, 625--706.

\bibitem[Hunt \& Carruthers(1990)]{HC}
{\sc Hunt, J.C.R. \& Carruthers, D.J.} 1990 Rapid distortion theory and the
  problems of turbulence. {\em J. Fluid Mech.\/} {\bf 212}, 497--532.

\bibitem[Klein(1995)]{K_1995}
{\sc Klein, A.} 1995 Characteristics of combustor diffuser. {\em Prog.
  Aerospace Sci.\/} {\bf 31}, 171--271.

\bibitem[Lavezzo {\em et~al.\/}(2010)Lavezzo, Soldati, Gerashchenko, Warhaft \&
  Collins]{LSGWC2010}
{\sc Lavezzo, V., Soldati, A., Gerashchenko, S., Warhaft, Z. \& Collins, L.R.}
  2010 On the role of gravity and shear on inertial particle accelerations in
  near-wall turbulence. {\em J. Fluid Mech.\/} {\bf 658}, 229--246.

\bibitem[Maxey \& Riley(1983)]{MR}
{\sc Maxey, M.R. \& Riley, J.R.} 1983 Equation of motion for a small rigid
  sphere in a nonuniform flow. {\em Phys. Fluids\/} {\bf 26}~(4), 883--889.

\bibitem[Ott \& Mann(2000)]{OM}
{\sc Ott, S. \& Mann, J.} 2000 An experimental investigation of the relative
  diffusion of particle pairs in three-dimensioinal turbulence. {\em J. Fluid.
  Mech.\/} {\bf 422}, 207--223.

\bibitem[Pope(2000)]{P_2000}
{\sc Pope, S.~B.} 2000 {\em Turbulent Flows\/}. Cambridge University Press.

\bibitem[Rogallo(1981)]{Rogallo_81}
{\sc Rogallo, R.~S.} 1981 Numerical experiments in homogeneous turbulence. {\em
  Tech. Rep.\/} 81835. NASA Tech. Mem.

\bibitem[Shaw(2003)]{S2003}
{\sc Shaw, R.} 2003 Particle turbulence interactions in atmospheric clouds.
  {\em Ann. Rev. Fluid. Mech.\/} {\bf 35}, 183--227.

\bibitem[Toschi \& Bodenschatz(2009)]{TB2009}
{\sc Toschi, F. \& Bodenschatz, E.} 2009 Lagrangian properties of particles in
  turbulence. {\em Ann. Rev. Fluid Mech.\/} {\bf 41}, 375--404.

\bibitem[Virant \& Dracos(1997)]{VD}
{\sc Virant, M. \& Dracos, T.} 1997 Ptv and its application on lagrangian
  motion. {\em Meas. Sci. Tech.\/} {\bf 8}, 1539--1552.

\bibitem[Voth {\em et~al.\/}(2002)Voth, Porta, Crawford, Alexander \&
  Bodenschatz]{VPCAB}
{\sc Voth, G.A., Porta, A.~La, Crawford, A.M., Alexander, J. \& Bodenschatz,
  E.} 2002 Measurements of particle accelerations in fully developed
  turbulence. {\em J. Fluid Mech\/} {\bf 469}, 121--160.

\bibitem[Warhaft(1980)]{W_1980}
{\sc Warhaft, Z.} 1980 An experimental study of the effect of uniform strain on
  thermal fluctuationsin grid–generated turbulence. {\em J. Fluid Mech.\/} {\bf
  99}, 545--–573.

\bibitem[Xu {\em et~al.\/}(2008)Xu, Ouelette \& Bodenschatz]{XOB_2008}
{\sc Xu, H., Ouelette, N.T. \& Bodenschatz, E.} 2008 Evolution of geometric
  structures in intense turbulence. {\em New J. Phys.\/} {\bf 10}, 013012.

\bibitem[Yeung \& Pope(1998)]{YP_1998}
{\sc Yeung, P.K. \& Pope, S.B.} 1998 Lagrangian statistics from direct
  numerical simulations of isotropic turbulence. {\em J. Comput. Phys.\/} {\bf
  79}, 373--416.

\end{thebibliography}
\end{document}